\documentstyle[aps,twocolumn,psfig,prl]{revtex}

\psfigurepath{figs}

\def \aoell{\overline{\ell}}
\def \opar{{\mathcal{L}}}
\def \order#1{{\mathcal{O}}(#1)}
\def \lanl#1{{\it cond-mat/#1}}
\def \recent {watts,BA,barrat,several,NW,newman2,corrlen,spreading}

\title{First-order transition in small-world networks}
\author{M.~Argollo de Menezes \and C.~F.~Moukarzel \and T.~J.~P.~Penna}

\address{Instituto de F\'{\i}sica, Universidade ~Federal Fluminense, CEP
  24210-340, Niter\'oi, RJ, Brazil}

\begin{document}

\maketitle

\begin{abstract}
  The small-world transition is a first-order transition at zero density $p$
  of shortcuts, whereby the normalized shortest-path distance $\opar =
  \aoell/L$ undergoes a discontinuity in the thermodynamic limit.  On finite
  systems the apparent transition is shifted by $\Delta p \sim L^{-d}$.
  Equivalently a ``persistence size'' $L^* \sim p^{-1/d}$ can be defined in
  connection with finite-size effects. Assuming $L^* \sim p^{-\tau}$, simple
  rescaling arguments imply that $\tau=1/d$. We confirm this result by
  extensive numerical simulation in one to four dimensions, and argue that
  $\tau=1/d$ implies that this transition is first-order.
\end{abstract}

A physically meaningful characterisation of disordered connection networks is
given by their average \emph{chemical distance} or shortest-path distance
$\aoell=1/N \sum_{j=1}^N <\ell_{ij}>$, where $\ell_{ij}$ is the minimum number
of links that must be traversed to join sites $i$ and $j$, and $<>$ means
average over disorder realizations.  If the network connectivity is
topologically $d$-dimensional, $\aoell$ is proportional to $L$, its linear
dimension. But on randomly bonded networks containing $L^d$ sites one has
$\aoell \sim \log(L)$ much shorter than $L$.  Small-world
networks~\cite{watts} interpolate between these two cases. They consist of a
regular $d$-dimensional lattice with a small fraction $p$ of long-range bonds,
or \emph{shortcuts}, per site (i.e. there is a total of $B=p L^d$ long-range
bonds on top of a regular lattice). These systems have received attention
recently~\cite{\recent}, following the observation~\cite{watts} that a very
small fraction of long-range bonds is enough to produce a transition to a
regime where $\aoell \sim \log(L)$.

Barth\'el\'emy and Amaral (in the following BA)~\cite{BA} recently showed that
for any nonzero $p$, the logarithmic regime is always reached if large enough
systems are considered. In other words there is a ``crossover size'' $L^*(p)$
such that for system size $L < L^*$, $\aoell(L,p)$ scales linearly with $L$,
while for $L>L^*$ $\aoell(L,p)$ grows as $\log L$. This behaviour is described
by the scaling form
\begin{equation}
\aoell(L,p) \sim L f(L/L^*(p))
\label{eq:scaling}
\end{equation}
with $f$ a scaling function such that $f(x) \to $ constant for $x << 1$ and $f
\to \log(x)/x$ for $x>>1$. BA however claim that this ``small world
transition'' is not a phase transition but a ``crossover phenomenon''. These
authors further propose that $L^*(p) \sim p^{-\tau}$ for $p \to 0$. By
numerical analysis on relatively small systems, they find $\tau \approx 2/3$
in one dimension and conclude that $\tau < 1$ provides evidence for the
importance of ``nonlinear effects''.

Some of their conclusions have been subject to criticism recently. Firstly
Barrat~\cite{barrat} stressed the obvious fact that $\tau$ cannot be smaller
than $1$ in one dimension, otherwise the total number of long-range bonds at
the ``crossover point'' would go to zero with system size.  Barrat confirmed
by numerical simulation on large systems that $\tau_{1d}=1$. Similar arguments
can be used to conclude that $\tau \geq 1/d$ in $d$ dimensions.

A second controversial point concerns the nature of the small-world
transition, which is the subject of this letter. Newman and Watts (thereafter
NW)~\cite{NW} recently suggested that the transition from logarithmic to
linear scaling of $\aoell$ is a normal \emph{second-order} phase transition at
$p=0$, i.e. one with a diverging characteristic length $\xi(p) \sim p^{-\nu}$,
and derive $\nu = 1/d$ by renormalisation-group (RG) analysis.
 
In this Letter we show that the small-world transition is a \emph{first-order
  transition} at $p=0$, where the ``order parameter'' $\opar(p)=\aoell/L$
undergoes a discontinuity in the thermodynamic limit.  Within this picture,
$L^* \sim p^{-\tau}$ in (\ref{eq:scaling}) is a \emph{persistence size}
associated with finite-size corrections. We derive $\tau=1/d$ by rescaling
arguments similar to those of NW and confirm this result by extensive
numerical simulation in one to four dimensions. The fact that $L^*\sim
p^{-1/d}$ implies that, on finite systems, there is a \emph{shift} or
\emph{broadening} of order $\Delta p \sim L^{-d}$ in the apparent transition
point.  In support of our claim for a first-order phase transition we recall
earlier work of Nienhuis and Nauenberg~\cite{NN} and of Fisher and
Berker~\cite{FB}, who first discussed the relationship between first-order
transitions and finite-size corrections of order $L^{-d}$.

Let us in the first place show that, if Eq.~(\ref{eq:scaling}) holds with $L^*
\sim p^{-\tau}$, then $\tau=1/d$. The first steps of our derivation are
similar to those in NW~\cite{NW}. Consider a decimation transformation with
rescaling parameter $b <<L$, i.e.  define ``blocks'' of linear size $b$, each
containing $b^d$ sites of the original lattice. After rescaling we are left
with a lattice of linear dimension $\widetilde{L}=L/b$.  Since all linear
dimensions are rescaled by $b$, one also has that $\widetilde{L^*}=L^*/b$.
Two blocks $I$ and $J$ in the decimated lattice are connected by a long-range
bond if and only if any pair of original sites $i \subset I$ and $j \subset J$
are connected by a long-range bond.  Since the total number $B$ of long-range
bonds is invariant under rescaling (because their typical length is $\sim L$,
much larger than $b$), one has $pL^d=\widetilde{p} \widetilde{L}^d$, and
therefore $\widetilde{p}=pb^d$.  After rescaling one must have
$\widetilde{L^*} \sim \widetilde{p}^{-\tau}$ and therefore $b =
L^*/\widetilde{L^*} = (p/\widetilde{p})^{-\tau}=b^{d\tau}$, which implies
$\tau=1/d$.

\begin{figure}
\centerline{\psfig{figure=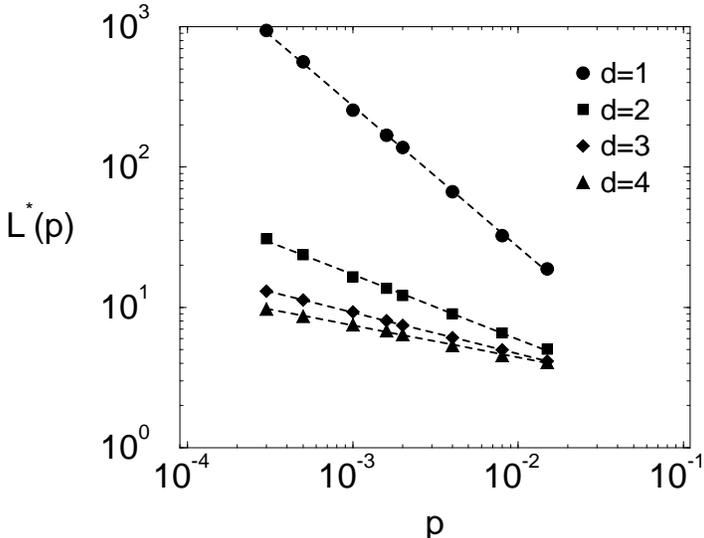,width=9cm}}
\caption{ {} Persistence size $L^*$ \emph{vs} $p$ in one to four
  dimensions. Our fits for $\tau$ such that $L^* \sim p^{-\tau}$ agree, within
  $5\%$ error in all cases, with $\tau=1/d$, the value predicted by rescaling
  arguments.  
  }
\label{fig:1}
\end{figure}

Our extensive numerical investigations confirm this theoretical prediction. We
used hypercubic lattices with helicoidal boundary conditions, of sizes up to
$250000$ lattice sites in one to four dimensions, to which random bonds were
added with probability $p$ per site. Since in this problem all bond costs are
unity, shortest-path distances $\ell_{ji}$ can be determined efficiently by
means of Breadth-First-Search (BFS)~\cite{algs}, whose time-complexity is
${\mathcal{O}}(N)$, and is simple to program.  We calculate $\aoell(L,p)$ by
averaging over $1000-2000$ disorder realizations for several lattice sizes and
$p$ values.  From Eq.~(\ref{eq:scaling}), the asymptotic derivative $\partial
\aoell / \partial \log{L}$ becomes proportional to $L^*$ as $L\to \infty$. We
thus fit a straight line to the $\aoell$ vs. $\log{L}$ curve, using the four
largest sizes simulated for each value of $p$, and obtain the values of
$L^*(p)$ displayed in Fig.~\ref{fig:1}.  Fitting these data we obtain
$\tau=1/d$ within $5 \%$ error in all cases.

%%%%%%%%%%%%%%%%%%%%%%%%%%%%%%%%%%%%%%%%%%%%%%%%%%%%%%%%%%%%%%%%%%%%%%
%%%%%%%%%%%%%%%%%%%%%%%%%%%%%%%%%%%%%%%%%%%%%%%%%%%%%%%%%%%%%%%%%%%%%%

Now consider the order parameter $\opar(p)=\lim_{L \to \infty}{\aoell/L}$,
which is zero for all $p>0$, but a (lattice-dependent) constant for $p=0$.
This suggests that it is reasonable to identify the small-world transition as
a discontinuous, or \emph{first-order} phase-transition at $p=0$.  Our
characterisation of the transition as first-order is however not based on the
discontinuity of $\opar$, but a consequence of having $\tau =1/d$ as we now
discuss.
  
As first shown by Nienhuis and Nauenberg~\cite{NN} in the context of the
Renormalisation Group, and later more generally by Fisher and
Berker~\cite{FB}, at a first-order fixed point one eigen-exponent must take
the value $y=d$ (this is necessary in order to have phase-coexistence) and
therefore finite-size corrections of order $L^{-d}$ can be expected.
Conversely, if $y=d$ then some first-derivative of the free-energy density
undergoes a discontinuity at the transition, i.e. $y=d$ implies that the
transition is first-order. The existence of an eigen-exponent $y=d$ gives rise
to finite-size corrections (e.g. the shift in the apparent transition point)
which are of order $L^{-d}$~\cite{FB}. We see then that finite-size
corrections of order $L^{-d}$ are a signature of first-order transitions.

For pedagogical reasons let us first review how finite-size corrections of
order $L^{-d}$ arise at a typical first-order transition, and later make the
analogy with small-world transitions explicit.  Consider an Ising-type
$d$-dimensional ferromagnet of linear size $L$ at very low temperature,
subject to an external magnetic field $h$.  Assume that a single spin (e.g.
the central one) is ``pinned down''.  This system undergoes a first-order
phase transition at $h=0$.  In the thermodynamic limit, the magnetisation is
$m(h) \approx +1$ for $h>0$, and $m(h) \approx -1$ for $h\leq0$. For $L$
finite and $h>0$, most of the spins will point up if the bulk energy $hL^d$
associated with the field is larger than the energy $J\gamma$ associated with
breaking the $\gamma$ bonds around the pinned spin.  Therefore on finite
systems of size $L$, the transition no longer happens at $h=0$ but at
$hL^d-J\gamma =0$, i.e.  it suffers a shift $\Delta h$ of order $L^{-d}$.
Equivalently, for each finite $h \to 0^+$ a \emph{persistence size} $L^*\sim
h^{-1/d}$ can be defined such that $m<0$ for $L<L^*$ , while $m>0$ for
$L>L^*$.

In the case of small-world systems, the order parameter $\opar(p)$ is a step
function at $p=0$ when $L \to \infty$. But for finite sizes $L$,
$\opar(p,L)=\aoell(p,L)/L$ is nonzero not only at $p=0$ but for all $p <
\Delta p \sim L^{-d}$, or equivalently for $ L < L^* = p^{-1/d}$ when $p$ is
fixed. The situation is clearly the same as in our previous example, i.e. the
transition suffers a shift of order $L^{-d}$ on finite systems. 

Thus $L^*$ is not a ``crossover length''~\cite{BA} but reflects the existence
of boundary-dependent finite-size corrections around a first-order transition.
We thus prefer to call $L^*$ ``persistence size''~\cite{FB}.

As discussed in the previous paragraphs, by showing that the finite-size
corrections exponent $\tau$ equals $1/d$ we have also demonstrated that the
small-world transition is first-order. But contrary to what is sometimes
assumed, the first-order character of a transition does not preclude the
existence of critical behaviour. Notice that $\mathcal{{L}}$ behaves in the
same way as the magnetisation density $m(T)$ of a one-dimensional Ising model,
or the spanning-cluster density $P_{\infty}(p)$ in a one-dimensional
percolation model. These two cases do display critical behaviour, and
constitute simple examples of \emph{first-order critical points}, respectively
at $T=0$ and $p=1$.  First-order critical points have been briefly discussed
by Fisher and Berker (FB)~\cite{FB}.

It is usual (though not entirely correct) to refer to 1d percolation or 1d
Ising as displaying ``second-order'' transitions.  The correct designation is
``first-order critical''.  Around a FOCP finite size corrections $\Delta p$
are of order $L^{-d}$ because this is true for any discontinuous transition.
Since at a critical point one also has $\Delta p \sim L^{-1/\nu}$ this in turn
means that $\nu=1/d$ for a FOCP~\cite{FB}.  Thus the fact that the
\emph{global} observable $\aoell$ for finite small-world networks is a
function of a single scaling variable $x=Lp^{1/d}$ is compatible both with a
non-critical first-order transition (in which case $x=L/L^*$ with $L^*$ a
persistence size) and with a first-order critical point (in this case
$x=L/\xi$ with $\xi \sim p^{-1/d}$ an internal characteristic length).  

Newman and Watts~\cite{NW,newman2} characterise the small-world transition as
a ``normal second-order'' phase transition, i.e. they assume critical
behaviour (``first-order critical'' would be the appropriate term if there
were critical behaviour).  As said, the finite size behaviour of $\aoell(L,p)$
cannot be used to distinguish between a normal first-order point and a FOCP. A
possible way to prove or disprove the existence of critical behaviour in
small-world networks is to look at \emph{bilocal observables}, and see whether
an internal correlation length can be identified that governs the physics of
the transition and diverges at $p_c$.  We study in this work the behaviour of
a particular bilocal observable: $\ell(r)$, the average shortest-path distance
between two points separated by an Euclidean distance $r$.  The basic question
we wish to answer is whether the behaviour of $\ell(r)$ in the thermodynamic
limit is dictated by a finite characteristic length $\xi(p)$ that diverges at
$p=0$.

%%%%%%%%%%%%%%%%%%%%%%%%%%%%%%%%%%%%%%%%%%%%%%%%%%%%%%%%%%%%%%%%%%%%%%
\begin{figure}[thpb]
\centerline{\psfig{figure=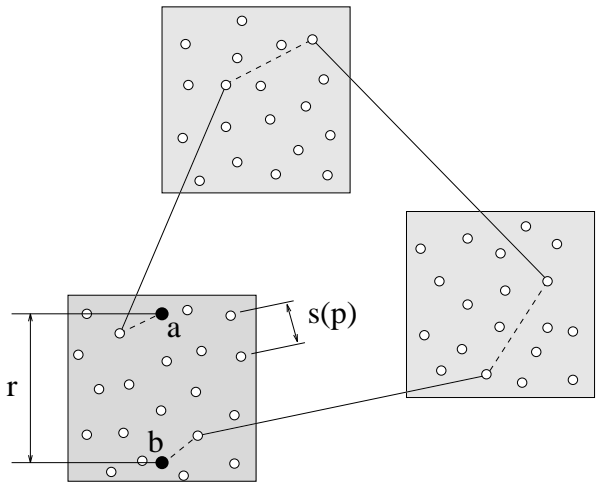,width=9cm,angle=270}}
\caption{ {} 
  We divide the system into blocks (shaded) of size $r$ (the Euclidean
  distance between $a$ and $b$) in order to estimate the average shortest-path
  distance $\ell(r)$ when $r$ is much larger than the mean separation
  $p^{-1/d}$ between shortcut-ends (white dots). By considering the distance
  travelled on the lattice (dashed lines) along any path from $a$ to $b$
  through shortcuts (black lines), one finds that $\ell(r)=r$ unless $r >
  p^{-1/d} \log L$ (see text).  }
\label{fig:2}
\end{figure}
%%%%%%%%%%%%%%%%%%%%%%%%%%%%%%%%%%%%%%%%%%%%%%%%%%%%%%%%%%%%%%%%%%%%%%

Newman and Watts~\cite{newman2} recently argued that the characteristic length
$\xi$ for the small-world transition is the mean separation $s(p)\sim
p^{-1/d}$ between shortcut-ends.  Their interpretation is consistent but they
apparently base this claim on the sole fact that the global observable
$\aoell$ is a function of $L/s(p)$ only. No analysis is done of any bilocal
observable, which we think is the only way to assess the existence of critical
behaviour.

We next show that $s(p)$ is \emph{not} a lengthscale dictating the behaviour of
shortest-path lengths $\ell(r)$.  We will find that a characteristic length
$r_c$ does in fact exist for $\ell(r)$, but that this characteristic length is
$r_c \sim s(p) \log L$, i.e. it diverges with system size, for all $p>0$.

Consider the shortest-path distance $\ell(r)$ between two points $a$ and $b$
separated by an Euclidean distance $r$. If $r<<s(p)=p^{-1/d}$ then clearly
$\ell(r)=r$. Let us then discuss the case $s(p)<< r << L$. Imagine dividing
the lattice into blocks of linear size $r$ as shown in Fig.~\ref{fig:2}. The
resulting rescaled system can be viewed as a random graph made up of
$\widetilde{N}=(L/r)^d$ nodes, randomly connected by a total of $B=pL^d$
long-range bonds. The average coordination number of a node (the number of
shortcuts connected to it) is $k \sim pr^d$.

Consider now a path from $a$ to $b$ through one or more shortcuts
(Fig.~\ref{fig:2}). In order to determine the ``cost'' of such path, we will
associate a cost zero with each shortcut, and only count the total distance
travelled on the Euclidean lattice \emph{within} each visited node (dashed
lines in Fig.~\ref{fig:2}).  There are typically no shortcuts with both ends
in the same block because $r<<L$. Thus a path $(a,b)$ through shortcuts must
contain two or more outgoing shortcuts and visit one or more extra nodes.
Thus we are looking for closed ``loops'' in the random graph, that start and
end in the original node (the one that contains $a$ and $b$). Now we know
that, starting from any given node, after $n$ shortcut-steps a total of $\sim
k^n$ nodes can be visited in average. A closed loop will exist if one of these
visited nodes is the original one. The probability for this to happen is
simply $k^n/\widetilde{N}$, which is negligible unless $n \sim \log
\widetilde{N}/\log pr^d$. Thus the shortest closed loop in the random graph of
blocks will typically contain $\order{\log \widetilde{N}/\log pr^d}$ or more
nodes.  The cost associated with visiting a node can be bounded to be at most
$\order{r}$ and at least $\order{s(p)}$.  Taking a conservative estimate let
us say it is $s(p)$. Then the lowest cost of going from $a$ to $b$ through
shortcuts is typically $\sim s(p) \log \widetilde{N}/\log pr^d$.  Thus it will
be ``cheaper'' to travel from $a$ to $b$ directly on the Euclidean lattice
than through shortcuts (and consequently $\ell(r) =r$) if $r <s(p) \log
Nr^{-d}/\log pr^d$.

\begin{figure}[thpb]
\centerline{\psfig{figure=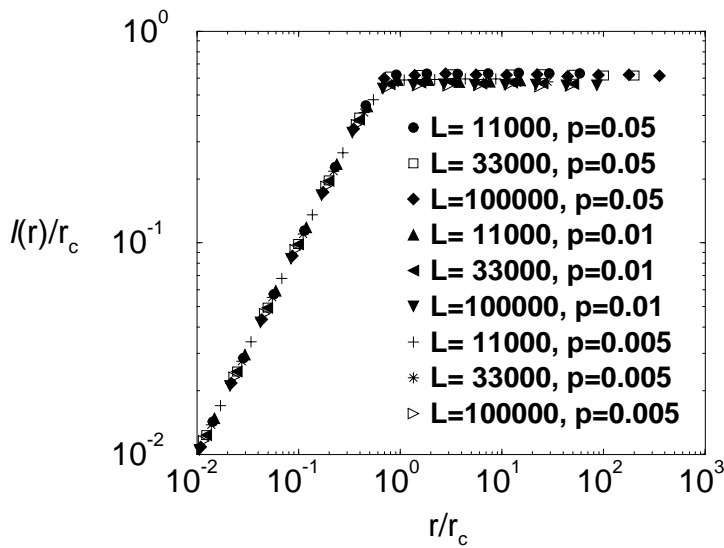,width=9cm,angle=270}}
\caption{ {} Shortest path distance $\ell(r)$ rescaled by
  $r_c$ vs. the rescaled Euclidean distance $r/r_c$, as measured on
  one-dimensional small-world lattices~\protect \cite{corrlen}. The
  characteristic length $r_c \sim p^{-1/d} \log L$ diverges with system size.}
\label{fig:3}
\end{figure}

This approximate scaling argument is confirmed in its main aspects by recent
analytical~\cite{spreading} and numerical~\cite{corrlen} studies (see
Fig.~\ref{fig:3}), which show that $\ell(r)$ behaves as

\begin{eqnarray}
\ell(r) \sim r ~~~\hbox{for}~~ r < r_c = p^{-1/d}\log(p^{1/d}L) \nonumber \\
 \sim r_c ~~\hbox{for}~~ r > r_c ~~~~~~~~~~~~~~~~~~~~~~~~~
\label{eq:ell}
\end{eqnarray}

In other words, the characteristic length for shortest-path distances
$\ell(r)$ is not $s(p)$ but $r_c$ above, which \emph{diverges} with $L$ for all
$p$.

%%%%%%%%%%%%%%%%%%%%%%%%%%%%%%%%%%%%%%%%%%%%%%%%%%%%%%%%%%%%%%%%%%%%%%
%%%%%%%%%%%%%%%%%%%%%%%%%%%%%%%%%%%%%%%%%%%%%%%%%%%%%%%%%%%%%%%%%%%%%%

In summary, we showed that the small-world transition is neither a ``crossover
phenomenon''~\cite{BA} nor a second-order phase transition~\cite{NW}, but a
first-order phase transition at $p=0$. Finite size corrections are dictated by
a ``persistence size'' $L^*(p)$ which behaves as $p^{-\tau}$~\cite{BA}.
Simple rescaling arguments similar to those used by Newman and Watts~\cite{NW}
show that $\tau=1/d$. This is confirmed by extensive numerical simulations in
one to four dimensions. We argue that having $\tau=1/d$ implies that the
transition at $p=0$ is \emph{first-order}~\cite{NN,FB}, and discuss the
possibility of further characterising this first-order point as a
\emph{first-order critical point}~\cite{FB}. This can only be done by looking
at bilocal observables and we choose to study $\ell(r)$. A relatively simple
scaling argument, confirmed by more rigorous studies~\cite{corrlen,spreading},
shows that the characteristic length $r_c$ dictating the behaviour of
$\ell(r)$ diverges with system size, i.e.  it is not possible to define a
finite $p$-dependent characteristic length $\xi(p)$ for $\ell(r)$ in the
thermodynamic limit. We suggest that $s(p)$ as defined by NW~\cite{newman2} is
an ``irrelevant'' internal length, i.e. one diverging at $p=0$ but playing no
role in determining shortest paths. A simple example of such an irrelevant
length is given by the following example: consider ferromagnet at $T << T_c,
h=0$ and with one site ``pinned down''. Now instead of a magnetic field $h$,
assume there is a density $p$ of randomly located sites whose spins are
``pinned-up''. As in our previous example, a first-order transition happens at
$p=0$ where the magnetization $m$ changes sign, and this transition is
\emph{clearly non-critical}.  In this ``tailored'' example, an internal
characteristic size $s(p) \sim p^{-1/d}$ (the average distance between pinning
sites) can be defined, which however plays no role in the physics of the
magnetic transition. This magnetic transition is a \emph{non-critical} first
order transition despite the fact that $s(p)$ diverges at $p=0$. We feel that
a similar phenomenon happens in the small-world transition since the average
distance between shortcut ends $s(p)$ diverges at $p=0$ but plays no role in
determining shortest paths.

\acknowledgments We thank M.~Barth\'el\'emy and L.A.N.~Amaral for sending us
their manuscript prior to publication, and P.~M.~C.~de Oliveira, D.~J.~Watts,
and M.~Newman for useful discussions and comments.  We acknowledge financial
support from CAPES, FAPERJ and CNPq.

\end{document}